\documentclass{paper}
\usepackage{graphicx}
\usepackage{float}
\usepackage{placeins}
\usepackage{array}
\usepackage[utf8]{inputenc}
\usepackage[backend=biber,
            style=ieee,
            sorting=none]{biblatex}
\usepackage{booktabs}
\usepackage{siunitx}
\usepackage{authblk}
\usepackage{amsmath,mathtools}
\usepackage{parskip}

\usepackage[colorinlistoftodos,prependcaption,textsize=small]{todonotes}

\usepackage[
    paperwidth=8.5in,
    paperheight=11in,
    left=1in,
    right=1in,
    top=0.8in,
    bottom=1in
]{geometry}

\title{Composition- and Ordering-Dependent Evolution of Simulated Kikuchi Patterns of Au-Ni Alloys}
\author[1]{Camila A. Teixera}
\author[1]{Lukas Berners}
\author[1]{Sandra Korte-Kerzel}
\affil[1]{Institute for Physical Metallurgy and Materials Physics, RWTH Aachen University, Aachen, Germany}
\author[2]{Ulrich Kerzel}
\affil[2]{Faculty of Georesources and Materials Engineering, RWTH Aachen University, Aachen, Germany}
\date{August 2026}
\begin{document}
\maketitle
\begin{abstract}
Understanding how Kikuchi patterns behave with subtle material variations is essential for developing machine learning (ML) based indexing methods for structurally and chemically complex cases, such as phases with potential sub-lattice order or (meta)stable defects with segregation (defect phases). Simulated Kikuchi patterns of the binary Au-Ni system were systematically analysed to investigate the effects of lattice parameter, chemical composition, partial site occupancy and ordering. The full compositional range from pure nickel to pure gold was considered, including a hypothetical ordered L$1_2$ Au$_3$Ni structure. Simulation parameters were optimised by comparison with experimental patterns. Normalized cross correlation showed limited sensibility to subtle differences between patterns. EMsoft simulation results revealed a systematic increase in mean intensity and more reflections contribute significantly as gold content increases. A surprising four-fold increase in mean intensity from 99\% gold to the pure gold sample highlighted limitations of partial site occupancy simulation by EMsoft. Difference maps showed enhanced normalised intensity along the \{111\} and \{200\} bands for the sample with 20\% gold compared to higher gold compositions. By isolating lattice parameter and chemical composition effects, chemical composition contributes predominantly to the mean intensity and strong reflections count, although the normalised intensity distribution was affected by both. Introducing L$1_2$ ordering increased the number of strong reflections and mean intensity, and redistributed normalised intensity along \{111\} and \{200\} bands and selected zone axis. Normalised intensity distribution and band width, are key descriptions to distinguish the Kikuchi patterns, which can be incorporated in future representation learning based indexing methods.
\end{abstract}

\section{Introduction}
\label{sec:intro}

Electron back-scatter diffraction (EBSD) is an important microstructure analysis technique that enables the study of crystallographic orientation and phase identification inside the scanning electron microscope (SEM) \cite{ebsd-book-schwartz-2009,WILKINSON2012-ebsd}. To obtain information from the microstructure of the analysed material, through their Kikuchi patterns, different indexing methods can be applied, such as the Hough transform \cite{krieger1996automatic} and Dictionary indexing (DI) \cite{chen-2015-DI}. Hough transform relies on band position identification and comparison of inter-planar angles with a pre-computed look-up table, while DI depends on a vast library of simulated patterns to find the best match through similarity metrics. Spherical indexing \cite{HIELSCHER-SI-2019,LENTHE-SI-2019}, as well as full pattern matching for orientation refinement \cite{winkelmann-2020}, have been proposed to improve indexing efficiency and overcome limitations of both the Hough transform and DI. However, analysis of specific changes within the lattice of a given phase, such as interplanar spacing of specific planes or site lattice occupancy \cite{Berners_arxiv_2026}, identification of ordered domains orientation \cite{MARTIN_2022} or even of defect phases with their atomic structural and chemical complexity \cite{Korte-Kerzel-def-phases-2022}, remain challenging with the currently available indexing methods. Changes in the Kikuchi pattern caused by defect phases might be too subtle to be recognized by conventional Hough transform \cite{SHARMA-hough-pattern-matching-2021}. Furthermore, computationally costly and time-consuming alternative methods highlight the demand for different approaches that would allow high throughput analysis of defect phases.

As machine learning (ML) continues to expand its capabilities, exploration of different ML models for indexing purposes became a generally attractive approach \cite{DING-cnn-ebsd-2020, Kaufmann-EBSDML-2020, Liu-EBSD-VAE-2025, Calvat2025}. The use of ML models to improve EBSD analysis of subtle changes within bulk phases as well as defect phases is particularly interesting, specifically implementation of representation learning \cite{Bengio-representation-learning-2013}. The goal is to use the neural network to learn meaningful features of the Kikuchi patterns (e.g. Kikuchi band width and positions, intensities, etc.), the so-called latent. Latents are low dimension vectors used by the neural network to describe a pattern, therefore, the learned features. Then, instead of relying on a vast library of simulated patterns for an image-to-image matching, the best match would be found through comparison between the learned latent and experimental patterns. This could enable a high throughput (defect) phase analysis, in contrast to the localised analysis currently performed, through transmission electron microscope (TEM) and atom probe tomography (APT) \cite{KorteKerzel-def-phases-2026}.

A key advantage of ML models compared to conventional indexing approaches is the possibility of incorporating different characteristics of a Kikuchi pattern, moving beyond simple band position or intensity variation detection. However, to leverage ML models best, it is crucial to understand how Kikuchi patterns behave as we modify aspects of the material. In this work, we focus mainly on comprehending how the change in chemical composition and ordering affects the Kikuchi pattern. Therefore, a systematic analysis of simulated Kikuchi patterns is carried out, using composition variations from pure nickel to pure gold, and a hypothetical ordered L$1_2$ structure with 75\% of gold and 25\% of nickel (Au$_3$Ni). The Au-Ni alloy is investigated in this work as it is of interest regarding defect phase studies, particularly given its tuneable miscibility gap and a rare ordering that could be present in the vicinity of defects \cite{NiAu-1998,NiAu-2005}. The main goal of this work is to observe the behaviour of the Kikuchi pattern as the chemical composition and degree of order change, providing essential groundwork to understanding what can be learned by a future representation learning model.

\section{Methods}
\label{sec:method}

\subsection{Experimental EBSD patterns acquisition}

To optimise simulation parameters, experimental patterns were acquired to allow comparison with simulated EBSD patterns. Pure nickel samples were metallographically prepared with grinding and polishing steps to obtain a deformation free surface, however, remaining deformation from the cutting process (done with pliers) was observed. The metallographic preparation consisted of grinding with SiC papers up to 4000 mesh, then polishing with 0.25~µm diamond suspension. To finalize sample preparation, an electropolishing step was performed with the Struers A2 electrolyte at 30~V with a flow rate of 8~m/s for 20~seconds, using the Lectropol-5 by Struers.
Then, the experimental patterns were acquired with a Clara Tescan Scanning Electron Microscope (SEM) with the EDAX Clarity Timepix 2 detector, using an acceleration voltage of 30 kV and beam current of 2.5 nA. All experimental patterns obtained were managed with the tools described in \cite{rejiba-openbis-2026}.
The sample was mounted on a \ang{70} pre-tilted holder and alignment was performed using the straight back side of the holder, however, due to the mounting of the Ni sample, a perfect parallel alignment might not be achieved. 

\subsection{EBSD pattern simulation}
EMsoft \cite{Callahan2013} was used for simulation of the master and screen patterns. EMsoft is a computational approach for dynamical EBSD pattern simulation based on Bloch waves theory with Monte Carlo (MC) simulation of backscattered electrons' (BSEs) energy, depth, and direction distribution \cite{Callahan2013}. Simulations were carried out in a high performance computing (HPC) cluster environment (CLAIX-2023) using the specifically developed EBSDmagus \cite{EBSDmagus2026} workflow manager. 

\begin{figure*}[hpbt]
   \centering
    \includegraphics[scale=0.6]{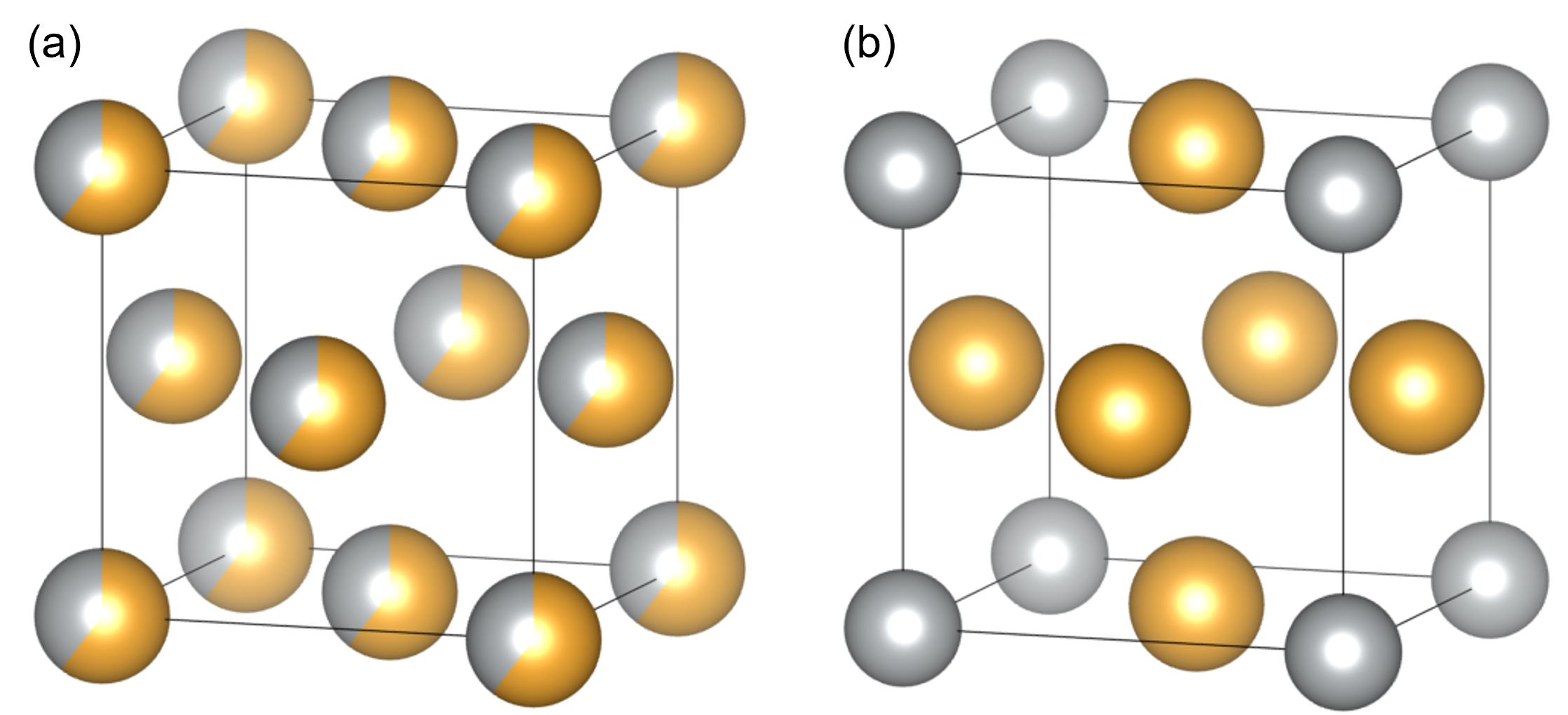}
    \caption{Crystal structure representation of (a) Au60Ni40 random state and (b) Au$_3$Ni ordered state (images produced using Vesta \cite{Momma-vesta-2011}).}
    \label{fig:fig_1}
\end{figure*}

To enable master and screen pattern simulation of partial site occupancy in Au-Ni samples (see in Fig.~\ref{fig:fig_1}(a)), and pure nickel and pure gold, crystallographic information files (CIF) were created and converted to xtal (format accepted by EMsoft). The samples' chemical composition varied from pure nickel to pure gold with 20\% increments of gold. Their lattice parameters were calculated based on Vegard‘s law \cite{Vegard192117}: 

\begin{equation}
\label{z-mean}
a_{\mathrm{alloy}} \;=\; (1 - x)\,a_{\mathrm{Au}} \;+\; x\,a_{\mathrm{Ni}},
\end{equation}

where $a_{Au}$ is the lattice parameter of gold (4.078~\AA) \cite{gold-lpr-Dutta1963} and $a_{Ni}$ is the lattice parameter of nickel (3.524~\AA) \cite{Ni-lprVoronin2016}. In this case, the contributions of both lattice parameter and variation of element content were jointly analysed. However, to understand their individual contributions, the lattice parameter and chemical composition were also investigated in a decoupled manner. Therefore, samples with a combination of constant lattice parameter (3.8010~\AA\ - Au50Ni50) and varying chemical composition (from Au20Ni80 to Au80Ni20), as well as fixed chemical composition (Au50Ni50) and changing the lattice parameter (from 3.6348~\AA\ to 3.9395~\AA) were simulated.

An ordered Au$_3$Ni structure was also studied as a hypothetical case, since it is not stable at room temperature \cite{NiAu-1998,NiAu-2005}. The hypothetical ordered state was chosen as a study case, since understanding how the Kikuchi bands behave due to ordering is also relevant for defect phase studies. Here, the element in the minority goes to the corners, thus nickel, and gold to the face centres (see Fig.~\ref{fig:fig_1}(b)). For comparison, an additional sample with unordered structure and the same chemical composition was simulated.

\begin{table}[htbp]
    \centering
    \caption{Main parameter values used for the Monte Carlo, master pattern and screen pattern simulation.}
    \label{tab:tab_1}
    \begin{tabular}{l l l}
        \toprule
        \multicolumn{3}{c}{Monte Carlo} \\      
        \midrule
        Parameter & Meaning & Value \\[2pt]            
        \midrule
        $\text{sigma}$         & sample tilt angle & $70$ \\
        $\text{numsx}$         & number of pixels along x-direction      & $501$ \\
        $\text{totnum\_el}$    & total number of incident electrons          & $2\,000\,000\,000$ \\
        $E_{\text{keV}}$       & incident beam energy     & $30\,$keV \\
        $E_{\text{histmin}}$   & minimum energy considered          & $29.5\,$keV \\
        $\text{depthstep}$     & depth step size                  & $1.0\,$nm \\
        $\text{depthmax}$      & maximum depth considered & $100\,$nm \\
        \toprule
        \multicolumn{3}{c}{Master pattern} \\    
        \midrule
        Parameter & Meaning & Value \\[2pt]            
        \midrule
        $d_{\min}$            & minimum inter‑planar spacing  & $0.02/0.05/0.07\,$nm \\
        $n_{\text{px}}$       & number of pixels along x-direction of the square master pattern \\ &(total number = 2*npx+1)       & $1000$ \\
        Bethe parameter       & cutoffs for strong, weak and ignored beams  \\
        $c_{1}$               & strong beam cutoff                  & $40$ \\
        $c_{2}$               & weak beam cutoff                 & $50$ \\
        $c_{3}$               & complete cutoff                  & $50$ \\
        $s_{\text{gdbdiff}}$  & double diffraction limit        & $1.0$ \\
        \toprule
        \multicolumn{3}{c}{Screen pattern} \\    
        \midrule
        Parameter & Meaning & Value \\[2pt]            
        \midrule
        $L$            & detector distance & $19\,489.3$ \\
        $\text{numsx}$ & number of grid points in *x* direction (screen) & $514$ \\
        $\text{numsy}$ & number of grid points in *y* direction (screen) & $514$ \\
        $x_{\text{pc}}$& x‑coordinate of the pattern centre (pixels) & $-5.397$ \\
        $y_{\text{pc}}$& y‑coordinate of the pattern centre (pixels) & $65.9462$ \\
        $\text{energy\_min}$ & lower bound energy range & $29.5\,$keV \\
        $\text{energy\_max}$ & upper bound energy range & $30\,$keV \\
        \bottomrule
    \end{tabular}
\end{table}

Initially, a comparison with experimental EBSD patterns was performed to optimise simulation parameters, such as the minimum interplanar spacing $d_{min}$ and the Bethe parameters (further explanation in the next section). In this case, different $d_{min}$ were simulated (0.02~nm, 0.05~nm and 0.07~nm) and the Bethe parameters were chosen based on the previous study by Wang and De Graef \cite{WANG201635}. The parameter values used for MC, master and screen pattern simulations are given in Tab.~\ref{tab:tab_1}. Further details on EMsoft parameters can be found in \cite{Jackson2019}. Note that, during the screen pattern simulation step, pattern centre, detector distance and Euler angles considered were dependent on the indexing (commercial Hough based method) information of the experimental patterns, to resemble the experimental pattern centre.

\subsection{EBSD master and screen pattern analyses}
The comparison between experimental and simulated pattern was performed with normalised cross correlation (NCC), typically used in the EBSD community for pattern structural similarity check \cite{nolze-mainstream-2017,winkelmann-ncc-2019,winkelmann-2020}. NCC calculations were done after three pre-processing steps that included (i) background division of the experimental pattern using a Gaussian blurred pattern ($\sigma = 30$). Then, (ii) $Z$-score normalisation, which includes subtraction of the mean and division by the standard deviation, and (iii) a Gaussian blur ($\sigma = 1.0$) step were applied to both experimental and simulated patterns. The NCC is computed as follows:
\begin{equation}
\label{NCC}
NCC =
\frac{\displaystyle\sum_{i=1}^{M}\sum_{j=1}^{N}
\bigl(I_{1}[i,j]-\bar I_{1}\bigr)\,
\bigl(I_{2}[i,j]-\bar I_{2}\bigr)}
{\displaystyle
\sqrt{\sum_{i,j}\bigl(I_{1}[i,j]-\bar I_{1}\bigr)^{2}}\;
\sqrt{\sum_{i,j}\bigl(I_{2}[i,j]-\bar I_{2}\bigr)^{2}}},
\end{equation}

where $I_{1}[i,j]$ and $I_{2}[i,j]$ are the pixel intensity of the two patterns respectively, and $\bar{I_{1}}$ and $\bar{I_{2}}$ are the mean values. The NCC takes on values between $-1 < NCC < 1$, where 1 corresponds to high similarity, 0 to no similarity at all, and -1 to inverse contrast between the analysed patterns \cite{NCC-2007}.

Band profile measurements, mean intensity calculations, and $Z$-score normalized intensity difference map analysis were performed to compare master and screen patterns. The band profile measurements were carried out after $Z$-score normalisation of the entire pattern and no background subtraction was performed. The applied normalisation step removes global intensity offsets while retaining features such as the band profile shape. For each analysed Kikuchi band, normalized intensity profiles were extracted along vectors perpendicular to the band and averaged over the band length. Relative changes in band width were qualitatively assessed from the edge-to-edge extent of the averaged intensity profile.
The mean intensity calculations were performed considering only non-zero intensity data points, as master patterns have a zero intensity black mask surrounding them, and final results were achieved by simple division of the non-zero intensity data by the number of data points. Finally, normalised difference intensity maps were calculated after $Z$-score normalisation of each master and screen patterns' intensity separately.

\section{Results}

\subsection{Optimizing simulation parameters}
\label{sec:res_exp_sim_comparison}
\subsubsection{Important parameters}
In EMsoft, the workflow for dynamical simulation of EBSD patterns consists of three consecutive stages: Monte Carlo (MC), master pattern, and screen pattern simulation \cite{Callahan2013}. In the MC stage, Monte Carlo techniques are applied to model the generation of backscattered electrons (BSEs), considering the energy and spatial distribution, resulting in a BSE energy distribution file \cite{Callahan2013}. In the master pattern stage, the MC simulation results are combined with the dynamical scattering calculations. The dynamical matrix is then solved, producing a master pattern that stores the intensity distribution over the orientation space \cite{Callahan2013}. In the final stage, screen patterns are extracted by interpolation over the master pattern for a given pattern centre, detector distance and crystallographic orientation (Euler angles or quaternions) \cite{Callahan2013}. A more detailed description of each stage can be found in \cite{Callahan2013}. In each step, there are important parameters that should be considered to replicate the EBSD patterns observed experimentally. In the following, we list our considerations for the Monte Carlo and master pattern simulation. For the screen patterns, the experimental data (after indexing using a commercial Hough based method) such as pattern centre, detector distance and Euler angles were used.

The MC simulation stage models the primary beam trajectory (through the sample) and energy as scattering events occur, recording the energy, depth and direction of the last scattering event as either a minimum energy is reached or the electron escapes \cite{EMsoftEBSD2017}. Thus, the incident beam energy, minimum energy and maximum depth considered for the calculations are crucial parameters for the MC simulation ($E_{\text{keV}}$, $E_{\text{histmin}}$ and $\text{depthmax}$ from Tab.~\ref{tab:tab_1} respectively). First, the energy of the incident beam is matched to the experimental acceleration voltage as the maximum beam energy, 30~keV. For the minimum beam energy, we considered a loss of 0.5~keV (29.5~keV), as the energy of electrons that contribute significantly to the diffraction pattern is not too far from the primary beam energy \cite{DEAL2008116, winkelmann-ncc-2019}. The maximum depth considered was kept at 100~nm, as applied for pure nickel simulation in previous works \cite{KURNIAWAN2021147,Jackson2019}. Depths beyond that are expected to only increase the computation time rather than account for electrons that will actually contribute to the diffraction pattern formation.

In the master pattern simulation stage, the main parameters to consider are the $d_{min}$ and Bethe parameters. Both parameters directly affect not only computation time but also the simulated diffraction process and, consequently, the final diffraction pattern. $d_{min}$ is the minimum $d_{spacing}$ considered for the calculations. $d_{spacing}$ is the inter-planar spacing and, therefore, $d_{min}$ sets a cut-off in the reciprocal space and directly impacts the number of reciprocal lattice points allowed (number of reflections) \cite{EMsoftEBSD2017}. They are inversely related, smaller $d_{min}$ values correspond to a larger cut-off in reciprocal space and more reflections are considered for the calculations. Experimental patterns may not resolve all the reflections observed in the simulation, due to factors such as loss of coherence, energy spread, and surface damage that can contribute to pattern blurring. Thus, some bands might not be distinguished experimentally. Therefore, different $d_{min}$ values of 0.02~nm, 0.05~nm and 0.07~nm, were simulated and the screen patterns were compared with the experimental patterns to observe which cut-off best reproduces the observed blurring (see Fig.~\ref{fig:fig_2}).

The Bethe potential first-order perturbation approach is used to reduce computation time by limiting the size of the dynamical matrix \cite{WANG201635}. The Bethe parameters select cut-off values that classify which reflections are strong, weak, or negligible, based on their excitation error and potential coupling \cite{WANG201635}. The strong reflections will be included in the dynamical matrix, the weak ones are treated perturbatively and the other ones are ignored. Values of strong ($C_{s}$) and weak ($C_{w}$) cut-off were chosen based on the findings by Wang and De Graef \cite{WANG201635} to maintain computational accuracy, therefore, $C_{s}$ = 40 and  $C_{w}$ = 50. Another parameter to consider could be the Debye–Waller factor \cite{Gao-debye-waller-1999} that takes into account atomic thermal vibrations, however, we assume that temperature influences in this case would be negligible.

\subsubsection{EBSD experimental and simulation comparison}

\begin{figure*}[hpbt]
   \centering
    \includegraphics[width=\textwidth]{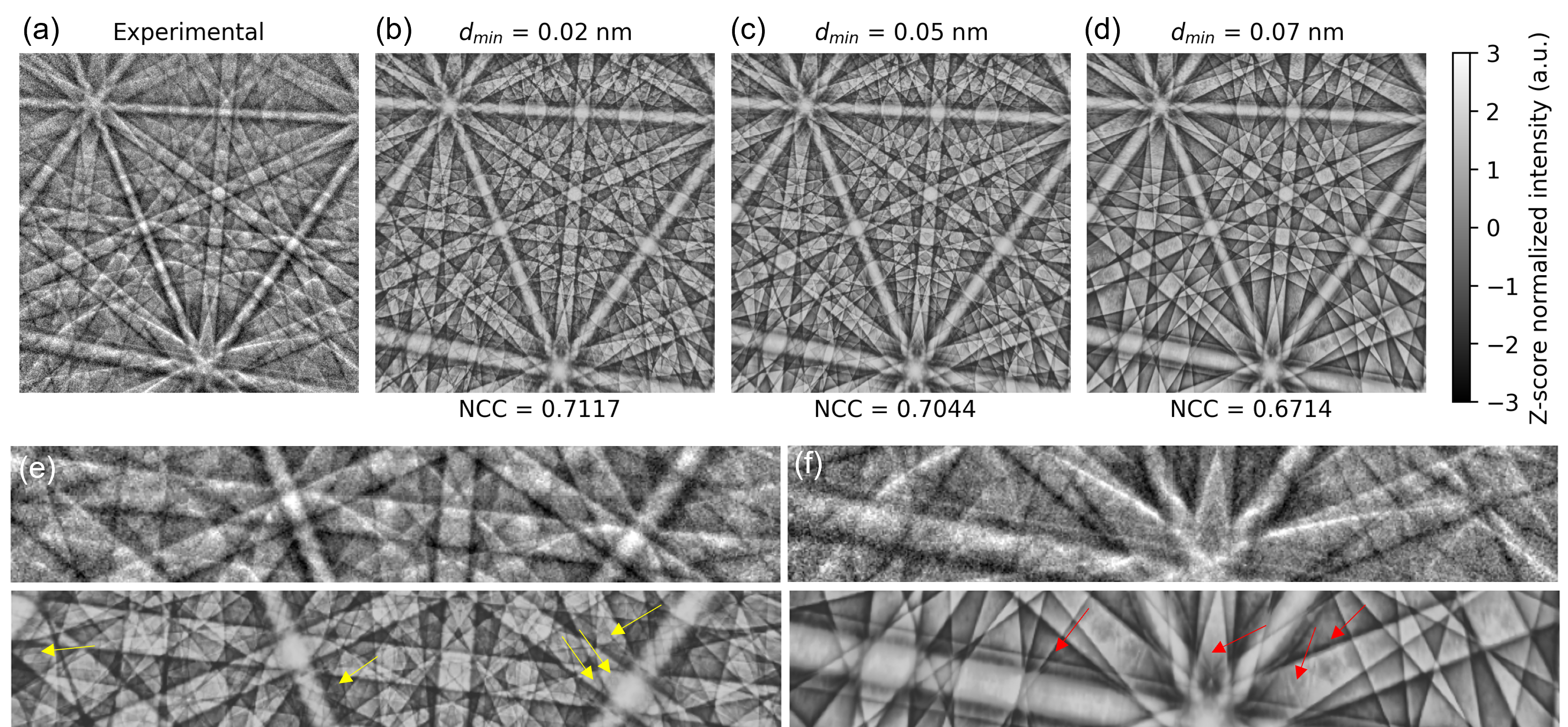}
    \caption{Pure nickel Kikuchi pattern obtained (a) experimentally with an acceleration voltage of 30~kV, and simulated patterns with a $d_{min}$ of (b) 0.02~nm, (c) 0.05~nm and (d)~0.07 nm, and respective NCC in comparison with the experimental pattern. In (e) a magnified section of the experimental and $d_{min}$ = 0.02~nm pattern for comparison, note that the yellow arrows highlight features that are not resolved experimentally. In (f) a magnified section of the experimental and $d_{min}$ = 0.07~nm pattern for comparison, here the red arrows highlights a few of the features missing in the simulation that are present in the experimental pattern.}
    \label{fig:fig_2}
\end{figure*}

After selecting EBSD simulation parameters, a comparison between simulated and experimental patterns is carried out to find the best agreement with the experiment. Fig.~\ref{fig:fig_2} shows the experimentally obtained pattern and the simulated patterns with different $d_{min}$, 0.02~nm (Fig.~\ref{fig:fig_2}(b)), 0.05~nm (Fig.~\ref{fig:fig_2}(c)) and 0.07~nm (Fig.~\ref{fig:fig_2}(d)) for pure nickel at 30~kV acceleration voltage and their respective NCC values in comparison to the experimental pattern (Fig.~\ref{fig:fig_2}(a)).

In Fig.~\ref{fig:fig_2}(b) through (d), as $d_{min}$ increases from 0.02~nm to 0.07~nm, progressively fewer features are observed. For instance, with a $d_{min}$ of 0.07~nm, higher order bands observed experimentally are not seen (see Fig.~\ref{fig:fig_2}(f) missing bands are highlighted with red arrows). In contrast, the lowest $d_{min}$ (0.02~nm) exhibits features which are not resolved experimentally, as seen in the magnified Fig.~\ref{fig:fig_2}(e) and missing features highlighted by yellow arrows. The change in $d_{min}$ directly affects the number of strong reflections included in the dynamical matrix, which decreased from 42 (0.02~nm), to 29 (0.05~nm) and finally 23 (0.07~nm). 

NCC metrics comparison was carried out to check structural similarity between the experimental and simulated patterns (see the NCC values below the patterns in Fig.~\ref{fig:fig_2}(b)-(d)). The results suggested that the $d_{min}$ of 0.02 nm is the most similar to the experimental pattern, even though there are clearly features which are not observed experimentally but exhibited in Fig.~\ref{fig:fig_2}(b) and (e). Comparing the NCC results between the simulated pattern, despite the subtle differences from patterns of $d_{min}$ = 0.02~nm and $d_{min}$ = 0.05~nm, a difference of only 0.007 was computed. Additionally, for the $d_{min}$ = 0.07~nm, which is the most inaccurate representation, NCC differences were around 0.04. Therefore, although NCC is a useful metric for comparison, it does not seem to detect subtle differences between the patterns, especially for the smaller $d_{min}$ values of 0.02~nm and 0.05~nm. In addition to the structural similarity metrics, computation time should also be considered, and for the $d_{min}$ = 0.02~nm that is approximately 3.5 times slower compared to $d_{min}$ = 0.05 nm.  In this case and specific experimental conditions, $d_{min}$ =0.05~nm seems to be the best match visually and computationally more efficient, even though the NCC is slight lower compared to 0.02~nm.

\begin{figure*}[hpbt]
    \centering
    \includegraphics[width=\textwidth, scale=0.67]{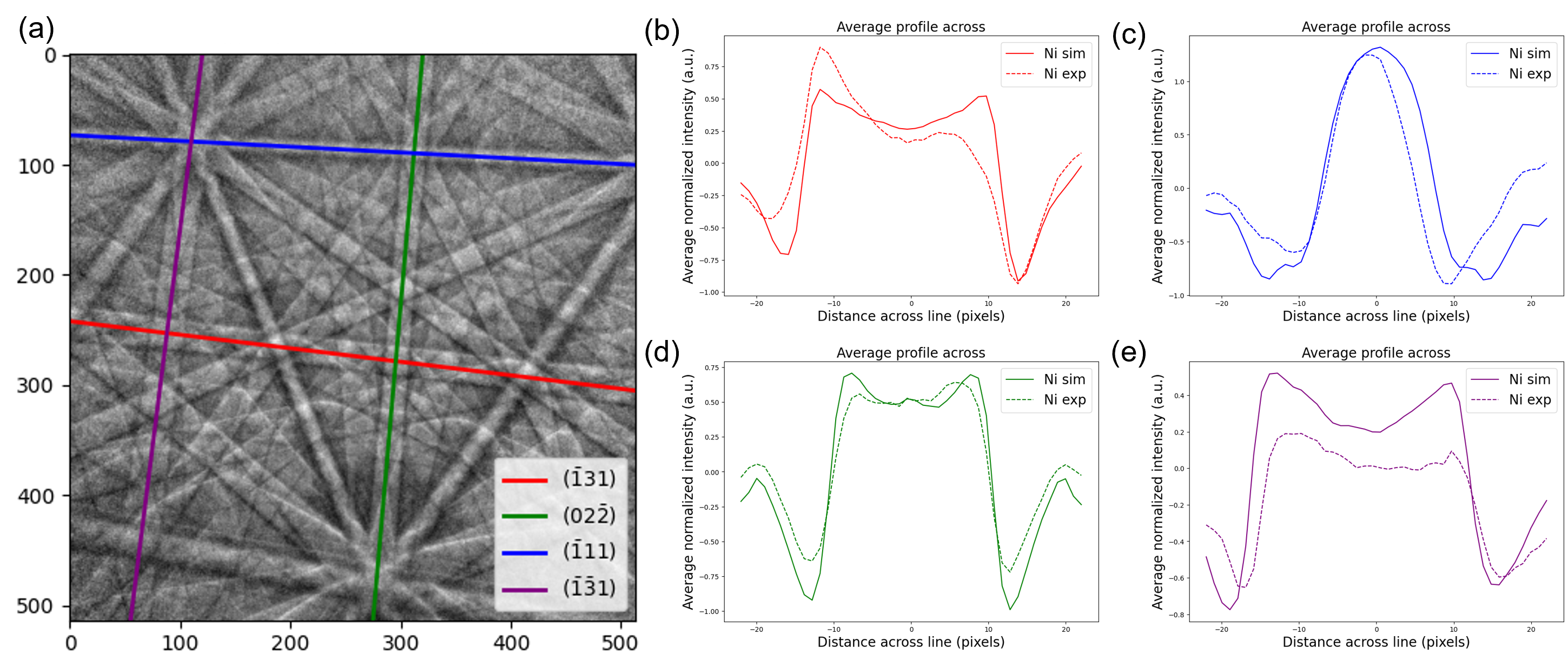}
    \caption{Experimental and simulated band profile comparison for pure nickel at 30~kV acceleration voltage. In (a) the path of the analysed bands taken from the simulated pattern and over-layered on the experimental pattern. In (b)-(e) band profiles for experimental (dashed lines) and simulated (full lines).}
    \label{fig:fig_3}
\end{figure*}

Line profile measurements were also carried out, to detect any misalignment between the experimental and simulated pattern, that could affect the calculated NCC values (Fig.~\ref{fig:fig_3}). In Fig.~\ref{fig:fig_3}(b)-(e), the line profile shows slight misalignment between the experimental and simulated patterns of pure nickel. Misalignments of around 2~pixels are noticed in the bands ($\bar{1}$31) and ($\bar{1}\bar{3}$1). The ($\bar{1}$11) band is broader than observed experimentally, also around 2~pixels. However, it is important to note that the screen patterns are simulated based on Euler angle and pattern centre values given after hough-based indexing of the experimental patterns. These differences could arise from sample preparation, as EBSD is a surface technique, any remaining deformation can affect the patterns. Additionally, any misalignment in detector distance (from the indexing procedure) could lead to differences in band width. Therefore, NCC values bellow 0.8 can be explained by these slight misalignments rather than any inaccurate simulation parameter choice. Based on the observed differences between the simulated and experimental patterns in this study, all further simulations were performed using a $d_{min}$ of 0.05~nm. This choice represents a suitable compromise for the present study, rather than an overall valid physically determined cut-off.

\FloatBarrier

\subsection{Changes in the simulated Kikuchi patterns}
\subsubsection{Mean intensity and number of strong reflections}
\label{sec:res_intensity_reflections}

\begin{figure}[hpbt]
    \centering
    \includegraphics[scale=0.6]{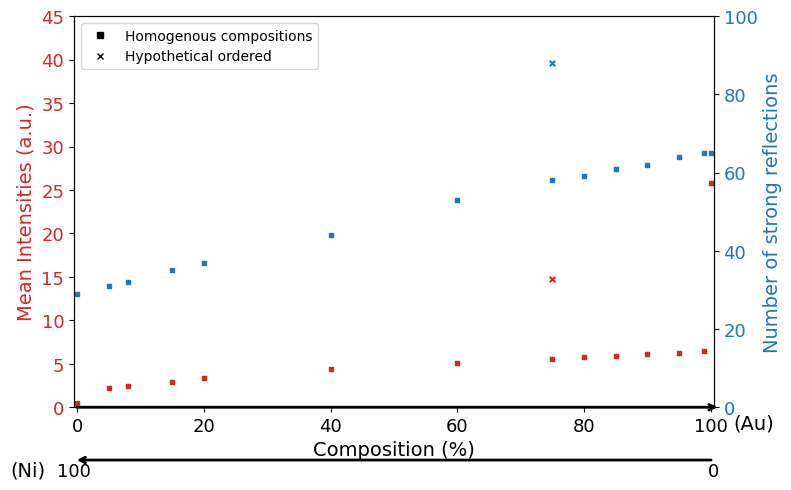}
    \caption{Calculated mean intensities (in red) of all simulated master patterns and number of strong reflections (in blue) that were included in the dynamical matrix, as functions of the chemical composition.}
    \label{fig:fig_4}
\end{figure}

To evaluate how the change in chemical composition (from pure nickel to pure gold) and in ordering affected the Kikuchi patterns' final intensity and dynamical matrix calculations, the mean intensity of each master pattern and the number of strong reflections included in the dynamical matrix were analysed (see Fig.~\ref{fig:fig_4}). It is clear that as gold is added to nickel, there is a systematic increase in mean intensity and number of strong reflections. However, the mean intensity does not follow the same relative increase as the number of strong reflections. The significant increase in mean intensity from pure nickel to 20\% gold (Au20Ni80) and from 80\% gold (Au80Ni20) to pure gold was surprising. This led to additional master patterns simulation of samples in between both compositions. Finally, the observed behaviour from pure nickel to 5\% gold (Au5Ni95) was relatively as expected compared to the expressive increase for compositions closer to pure gold. This is particularly noticeable from the 99\% gold (Au99Ni1) to the pure gold sample, as a four-fold mean intensity increase is observed. For the partial site occupancy samples, the raw mean intensity calculated does not seem to reflect the physical reality, specially considering two samples which are essentially the same. For the ordered sample case, Au$_3$Ni, results were as expected, with a noticeable increase in number of strong reflections as well as a higher mean intensity, compared to the unordered Au75Ni25 (see Fig.~\ref{fig:fig_4}).

\subsubsection{Variation of chemical composition}
\label{sec:res_composition}
\begin{figure*}[hpbt]
    \centering
    \includegraphics[width=\textwidth, scale=0.67]{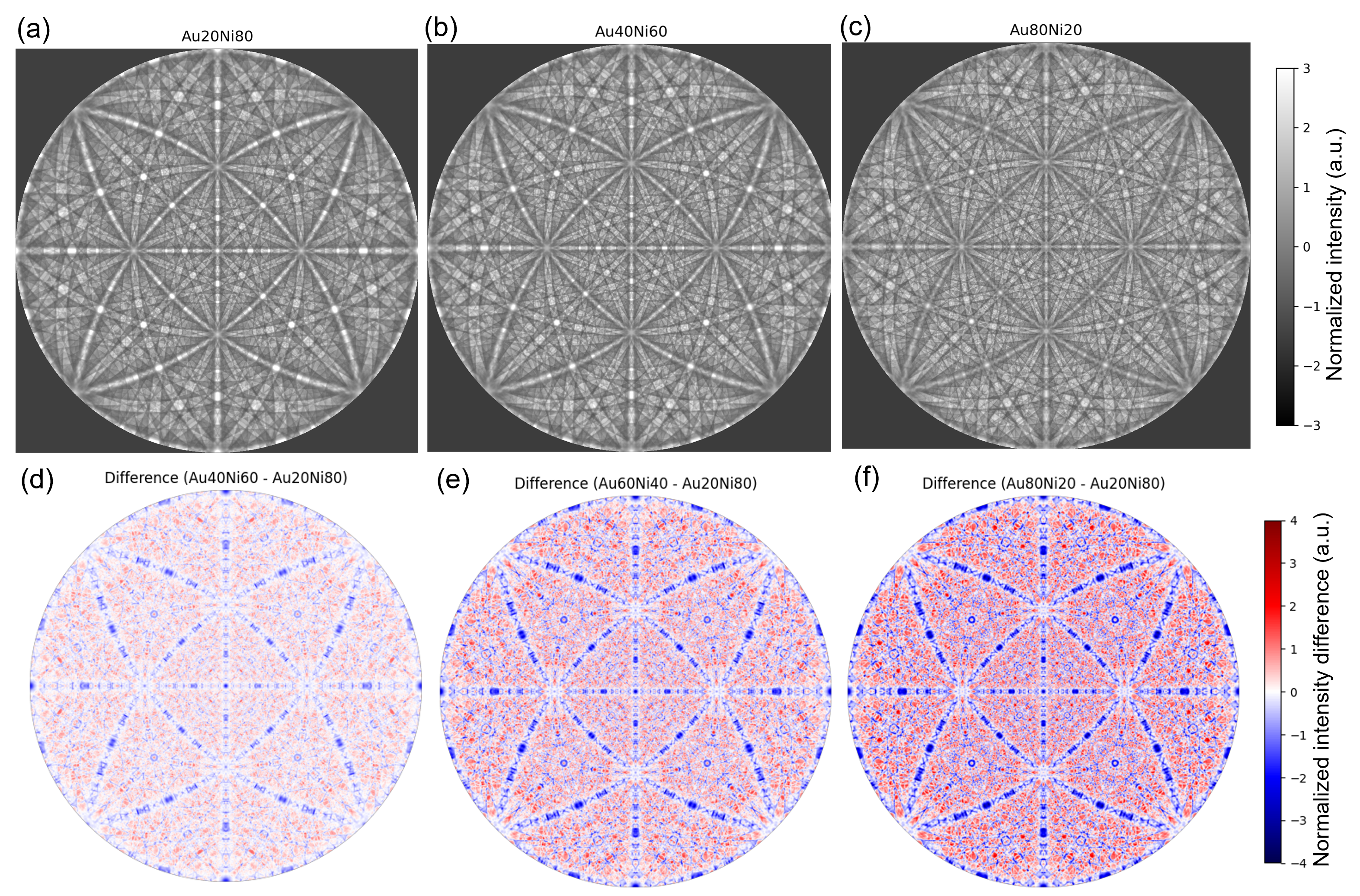}
    \caption{$Z$-score normalised master patterns of (a) Au20Ni80, (b) Au40Ni60 and (c) Au80Ni20. (d)-(f) the normalised intensity distribution difference between Au20Ni80 and (d) Au40Ni60, (e) Au60Ni40, and (f) Au80Ni20. Highlighted in blue are the regions in which Au20Ni80 has a higher normalised intensity and in red where the higher Au content samples have a higher intensity.}
    \label{fig:fig_5}
\end{figure*}

Changes in chemical composition can result in variations in the normalised intensity distribution in Kikuchi patterns. These variations can be observed through difference maps (see Fig.~\ref{fig:fig_5}), as it highlights the difference in the normalised intensity distribution, displaying the regions with enhanced normalised intensity for one chemical composition relative to the other. Comparing the simulated master patterns for samples Au20Ni80, Au40Ni60 and Au80Ni20, seen in Fig.~\ref{fig:fig_5}(a)-(c), only subtle differences in intensity can be observed and slightly broader band width for the lower gold content sample, given its smaller lattice parameter. The normalised difference maps, in Fig.~\ref{fig:fig_5}(d)-(f), show a higher normalised intensity for the \{111\} and \{002\} bands in the sample with 20\% gold, while in red the higher intensity value in the pattern for samples with higher gold content is observed. As gold content increases and no substantial change in structure actually occurs, the difference maps are similar and the main observable change is in the red and blue intensity, as their normalised intensity difference becomes more pronounced. 

\begin{figure*}[hpbt]
    \centering
    \includegraphics[scale=0.65]{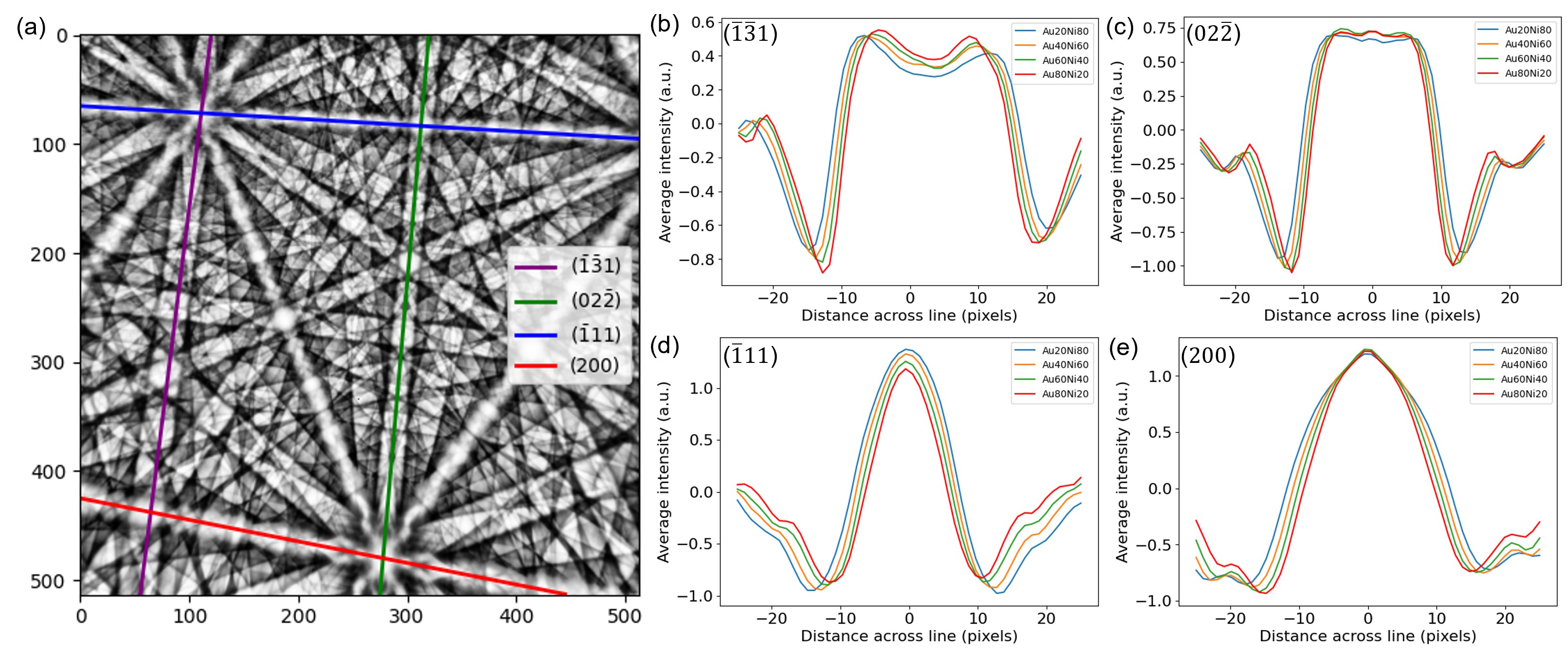}
    \caption{Band profile comparison for different chemical compositions of Au-Ni samples at 30~kV acceleration voltage. In (a) the path of the analysed bands over-layered on the top of the Kikuchi pattern and in (b)-(e) band profiles of the simulated compositions: Au20Ni80 (in blue), Au40Ni60 (in yellow), Au60Ni40 (in green) and Au80Ni20 (in red).}
    \label{fig:fig_6}
\end{figure*}

The band width is inversely proportional to the lattice parameter, therefore a smaller lattice parameter leads to broader Kikuchi band widths \cite{lpr-saowadee-2017,Nolze-lattice-paramenter-ebsd-2023}. Consequently, as chemical composition changes, the band width will be affected. To analyse the variations, band profile measurements were conducted and a comparison between the different chemical compositions is shown in Fig.~\ref{fig:fig_6}, with the respective Kikuchi pattern (Fig.~\ref{fig:fig_6}(a)) and paths of the band profile measured. As expected, in the band profile, the change in band width is clear. As lattice parameter increases, with addition of gold, the decrease in band width can be noticed. In red, the lines (Fig.~\ref{fig:fig_6}(b)-(d)) that represent the Au80Ni20 sample exhibit the narrowest band width and in blue, the lines that represent the Au20Ni80 sample, exhibit the broadest band width. Besides the band width becoming narrower as gold content increased, there were no significant changes in band shape for the analysed conditions.

\FloatBarrier

\subsubsection{Isolated effects: lattice parameter and chemical composition}
\label{sec:res_lpr_chem-comp}

\begin{figure*}[hpbt]
    \centering
    \includegraphics[width=\textwidth]{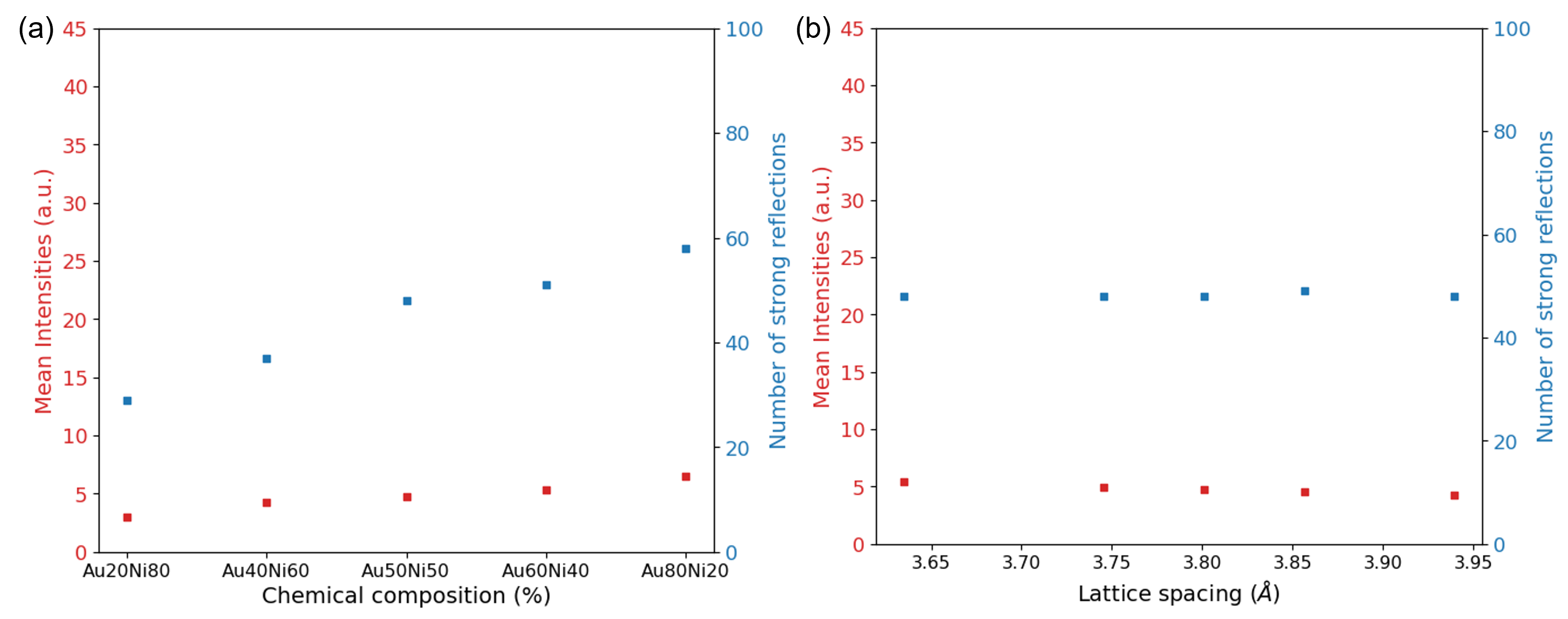}
    \caption{Mean intensities and number of strong reflections for the simulated master patterns with isolated changes in: (a) chemical composition and (b) lattice parameter.}
    \label{fig:fig_7}
\end{figure*}

\begin{figure*}[hpbt]
    \centering
    \includegraphics[width=\textwidth]{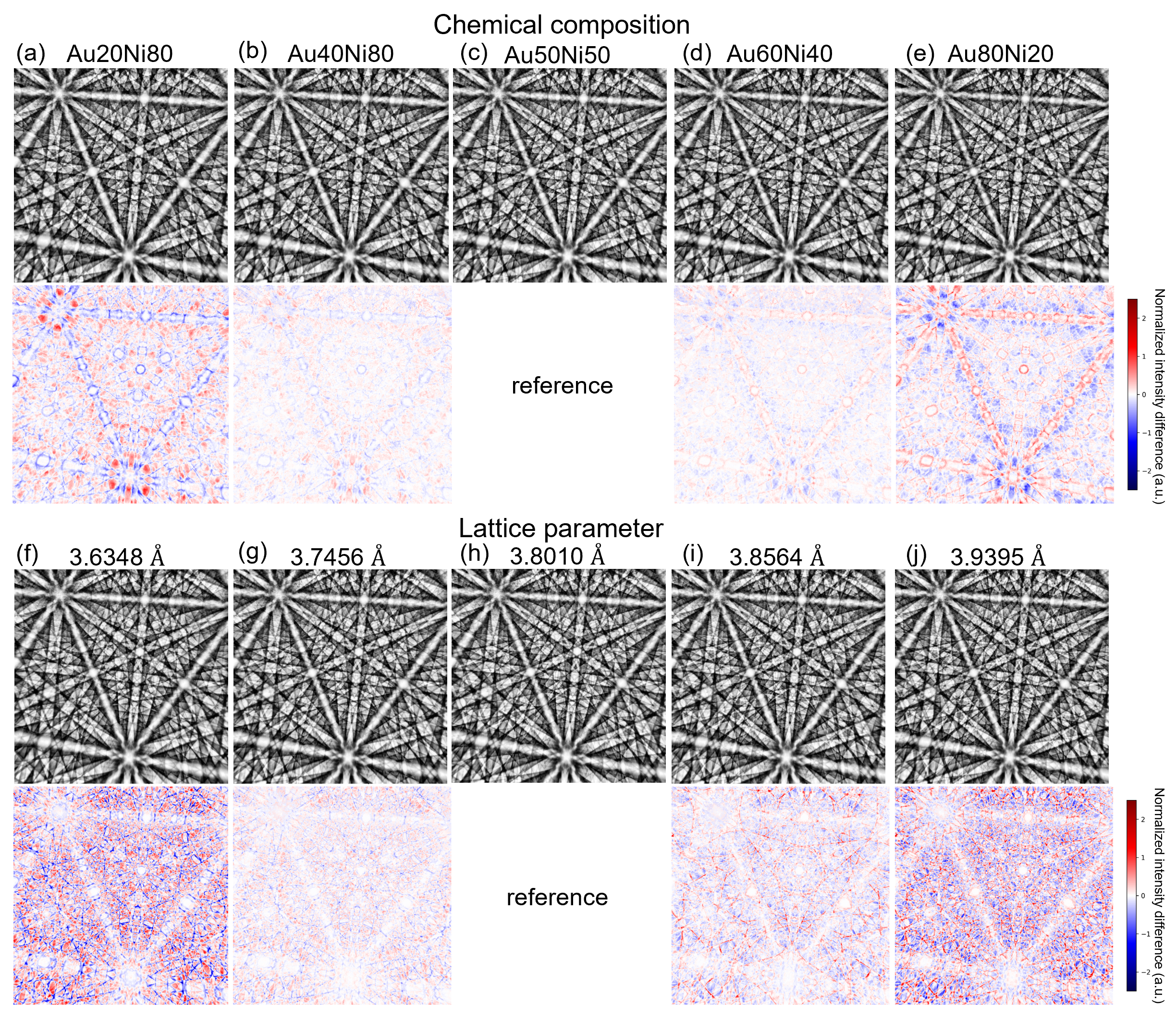}
    \caption{Screen patterns and normalised intensity difference maps comparing samples with different chemical composition (a) Au20Ni80, (b) Au40Ni60, (d) Au60Ni40 and (e) Au80Ni20, and samples with different lattice parameter (f) 3.6348~\AA, (g) 3.7456~\AA, (i) 3.8564~\AA\ and (j) 3.9395~\AA, with the reference sample Au50Ni50 with lattice parameter of 3.8010~\AA\ shown in (c) and (h).}
    \label{fig:fig_8}
\end{figure*}

\begin{figure*}[hpbt]
    \centering
    \includegraphics[width=\textwidth]{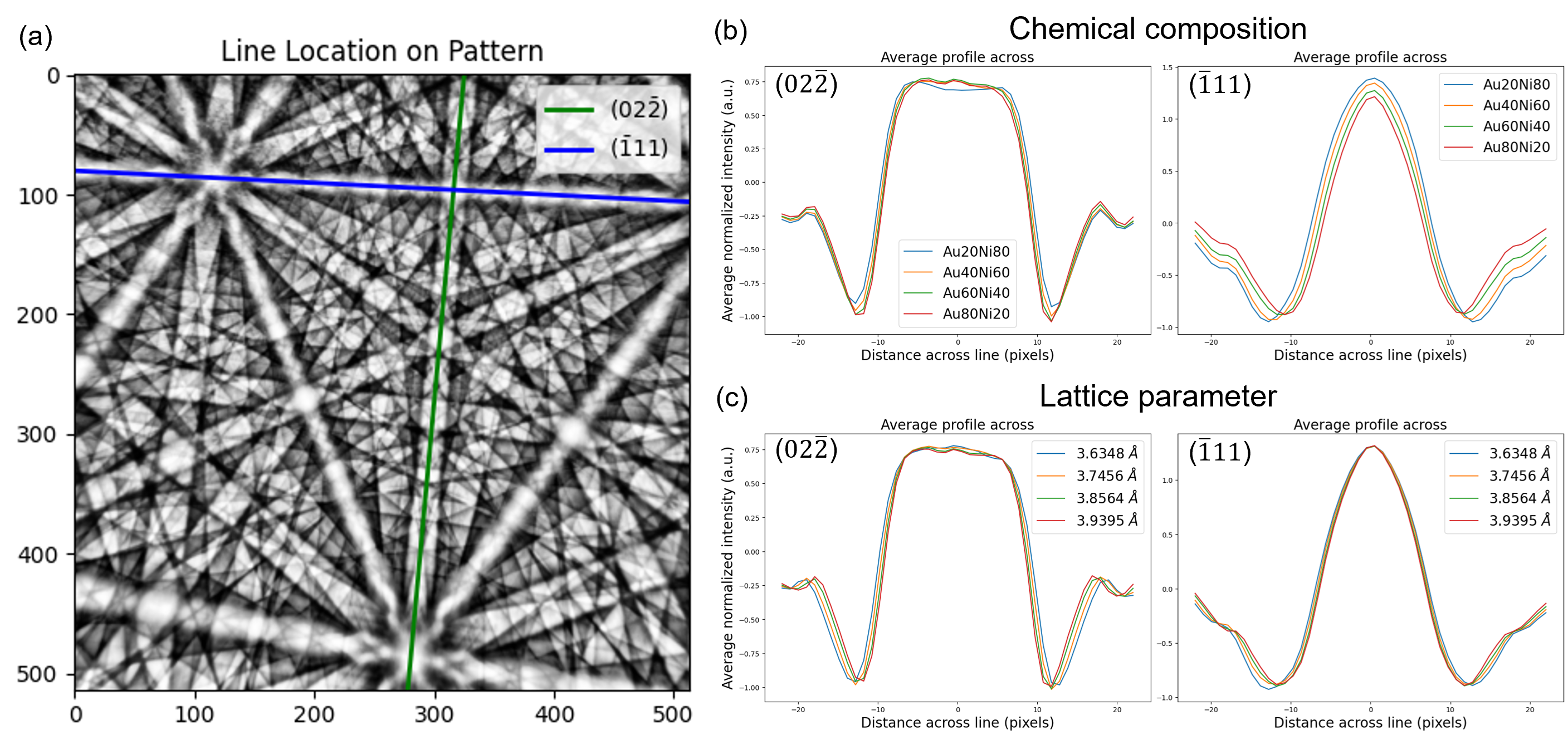}
    \caption{Band profile measurement comparison for (02$\bar{2}$) and ($\bar{1}$11) bands. In (a) the path of the analysed bands over-layered on top of the Kikuchi pattern and the isolated effect of (b) chemical compositions and (c) lattice parameter are shown.}
    \label{fig:fig_9}
\end{figure*}

In the previous section (\ref{sec:res_composition}), we have shown the combined effect of the chemical composition and lattice parameter. However, to understand their isolated contributions to the changes observed in the Kikuchi patterns, these effects can be studied individually through simulation. Therefore, samples with a constant lattice parameter (3.8010~\AA) but different chemical composition, as well as constant chemical composition (Au50Ni50) and varied lattice parameters were simulated. To observe these individual effects, mean intensity and number of strong reflections plot (Fig~\ref{fig:fig_7}), normalised intensity difference maps (Fig~\ref{fig:fig_8}) and band profile measurements (Fig.~\ref{fig:fig_9}) were performed.

Fig.~\ref{fig:fig_7} highlights the impact of the chemical composition variation on the mean intensity and number of strong reflections (Fig.~\ref{fig:fig_7}(a)) compared to lattice parameter (Fig~\ref{fig:fig_7}(b)). The mean intensity more than doubled as chemical composition varied from 20\% gold to 80\% gold content. The number of strong reflections accompanied this increase, moving from 29 to 58. While changes in the lattice parameter led to around 21\% reduction in mean intensity as lattice parameter was increased in 0.3047~\AA. Additionally, the number of strong reflections was not affected by the lattice parameter variation, with exception of the 3.8564~\AA\ sample with 49 strong reflections, a constant 48 strong reflections were included in the dynamical matrix.

The difference maps also display the distinct effect of lattice parameter and chemical composition in the normalised intensity distribution (see Fig. ~\ref{fig:fig_8}). In Fig.~\ref{fig:fig_8}(a)-(e), the blue regions highlight the higher normalised intensity for the different chemical composition, from 20\% gold (Au20Ni80) to 80\% gold (Au80Ni20), while in red the higher normalised intensity for the reference sample of 50\% gold (Au50Ni50). A higher difference in normalised intensity was observed predominantly in the zone axis and the \{111\} and \{002\} bands. In Fig.~\ref{fig:fig_8}(f)-(j), the blue regions highlight the higher normalised intensity for the different lattice parameters, from 3.6348~\AA\ to 3.9395~\AA, and in red the higher normalised intensity for the reference sample, 3.8010~\AA. The differences in normalised intensities are spread throughout the entire Kikuchi pattern with no predominant region. In the \{111\} and \{002\} bands and near zone axis, however, the normalised intensity seems to be more affected by the change in chemical composition, rather than lattice parameter.

The isolated effects of chemical composition and lattice parameter on the band profile were also investigated and the measurements are shown in Fig.~\ref{fig:fig_9}. The Kikuchi pattern in Fig.~\ref{fig:fig_9}(a) displays the paths and in Fig.~\ref{fig:fig_9}(b) and (c) the band profiles. A slight change in normalised intensity as chemical composition is varied, as well as a decrease in band width from 20\% gold (Au20Ni80) to 80\% gold (Au80Ni20) was observed, while the peaks position remained the same. Small changes in band width are also observed as lattice parameter were varied (see Fig.~\ref{fig:fig_9}(c)), decreasing with increase in lattice parameter, and a slight shift in the higher order band position are observed (see (02$\bar{2}$) band profile Fig.~\ref{fig:fig_9}(c)). Additionally, no changes in normalised intensity were observed.   

\subsubsection{Variation of degree of order }
\label{sec:res_ordering}

\begin{figure*}[hpbt]
    \centering
    \includegraphics[width=\textwidth, scale=0.67]{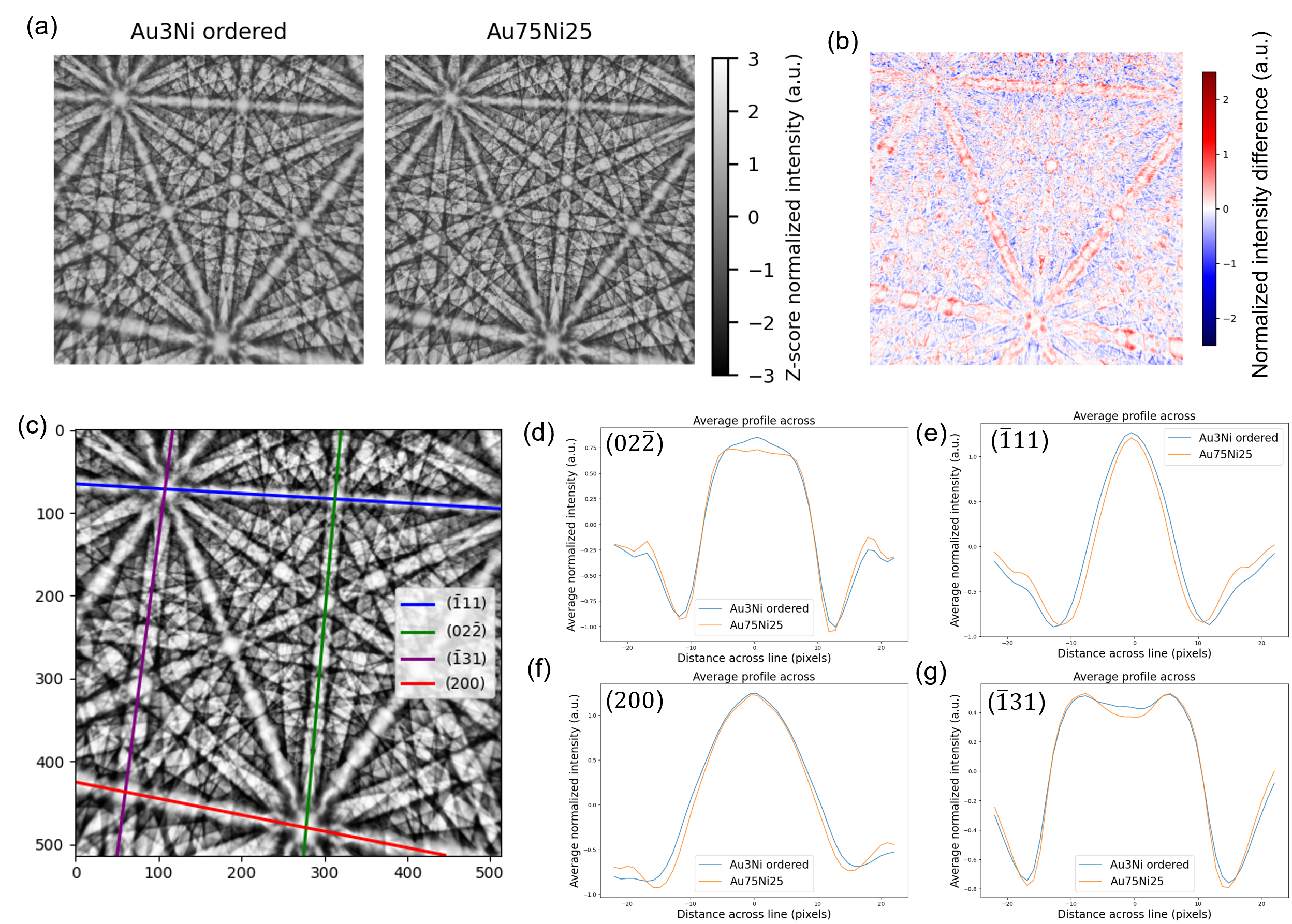}
    \caption{$Z$-score normalised screen patterns (a) for the Au$_3$Ni ordered state and unordered Au75Ni25. In (b) the normalised intensity distribution difference between the ordered Au$_3$Ni - unordered Au75Ni25. Highlighted in red are the regions in which the ordered state (Au$_3$Ni) has a higher normalised intensity and in red the regions the unordered state (Au75Ni25) has a higher intensity. In (c) the of the analysed bands over-layered in top of the Au$_3$Ni Kikuchi pattern and the line profile comparison of bands: (d) (02$\bar{2}$), (e) ($\bar{1}\bar{1}$1), (f) (200) and (g) ($\bar{1}$31).}
    \label{fig:fig_10}
\end{figure*}

Another important aspect to investigate is how the Kikuchi bands behave as ordering is introduced. Therefore, a comparison between the ordered sample, Au$_3$Ni, and an unordered sample with the same chemical composition, Au75Ni25, was carried out through a $Z$-score normalised intensity difference map (Fig.~\ref{fig:fig_10}(b)) and line profile measurements (Fig.~\ref{fig:fig_10}(d)-(g)). In Fig.~\ref{fig:fig_10}(a) shows the screen patterns for the ordered Au$_3$Ni and unordered Au75Ni25. Some subtle differences in terms of band width, for instance the (200) band seems slightly wider, as well as the zone axis, which appear blurred, are observed for the ordered structure in comparison to the unordered one. The normalised intensity difference map shows a clear redistribution of intensity predominantly along \{111\} and \{002\} bands, as well as the zone axis, for the ordered samples highlighted in red (see Fig.~\ref{fig:fig_10}(b)). In blue are the regions where the unordered Au75Ni25 normalised intensity prevails. Line profile measurements (shown in Fig.~\ref{fig:fig_10}(d)-(g)) exhibited slightly narrower band widths for Au75Ni25, compared to the Au$_3$Ni sample. In addition to that, we observed a peak rather than flattening in the (02$\bar{2}$) band (Fig.~\ref{fig:fig_10}(d)). The observed peak in the (02$\bar{2}$) band could be related to dynamical diffraction effects \cite{winkelmann2021}, rather than the presence of an overlapping superlattice reflection, such as (01$\bar{1}$). No significant change in band shape was observed for the other analysed bands.

\section{Discussion}
\label{sec:discussion}
\subsection{Increase in mean intensity and number of strong reflections}
\subsubsection{Chemical composition effect}
The systematic increase in mean intensity and number of strong reflections included in the dynamical matrix (Fig.~\ref{fig:fig_4}) as gold content increases was consistent with expectations. With each 20\% increment of gold, a high atomic number ($Z$) element, the mean atomic number ($\bar{Z}$) of each sample analysed increases (see Tab.~\ref{tab:tab_2}) and this impacts the scattering factor. With a higher $\bar{Z}$, there is an increase in the scattering factor, thus modifying the structure factors \cite{WINKELMANN-many-beams-2007}. The higher the structural factor, the stronger the electron scatter. This not only reflects in the progressive increase in the number of strong reflections, but also in the mean intensity, leading to the increase in scattering intensity observed \cite{nolze-mainstream-2017}.

Although the increase in mean intensity occurs as expected, the magnitude of this increase is surprising. The mean intensity of the pure gold sample was calculated as four times higher than the 99\% gold sample. When comparing two samples that have almost the same chemical composition, only a 1\% difference in gold/nickel, the assumption would be highly similar results. This was only partially observed by the same number of strong reflections computed, both with 65 strong reflections (as seen in Fig.~\ref{fig:fig_4}). However, in Tab.~\ref{tab:tab_2} we display two parameters extracted from the Monte Carlo and master pattern output files of each sample: $\bar{Z}$ and number of atoms in the unit cell. It is interesting to note that there is a difference in the number of atoms considered in the unit cell for the partial site occupancy samples, compared with pure nickel and pure gold, as well as the ordered sample, Au$_3$Ni. Despite partial site occupancy being correctly described in the crystal structure file (as observed in Fig.~\ref{fig:fig_1}(b)), EMsoft interprets a unit cell with 8 atoms rather than 4. Additionally, the calculation of $\bar{Z}$ does not seem to consider the partial site occupancy weight. This is evidenced by the same $\bar{Z} = 53.5$ for every partial site occupancy sample, even though there should be a difference based on the concentration of each element. The partial site occupancy the $\bar{Z}$ can be estimated as \cite{nolze-mainstream-2017}:

\begin{equation}
\label{z-mean}
\bar{Z} = \sum_{i=1}^{n} c_i Z_i
\end{equation}

where $c_i$ is the concentration of each element and $Z_i$ is its atomic number. This is correctly estimated as displayed on the Calculated $\bar{Z}$ column (see Tab.~\ref{tab:tab_2}). 

\begin{table}[htbp]
    \centering
    \caption{$\bar{Z}$ and number of atoms in the unit cell extracted from the EMsoft's output file and $\bar{Z}$ calculated for each sample.}
    \label{tab:tab_2}
    \begin{tabular}{llll} 
        \toprule
        Sample                & {EMsoft $\bar{Z}$} & {Calculated $\bar{Z}$} & {N.\ of atoms in the unit cell} \\
        \midrule
        Ni                    & 28        & -      & 4 \\
        Au20Ni80    & 53.5      & 38.2   & 8 \\
        Au40Ni60    & 53.5      & 48.4   & 8 \\
        Au60Ni40    & 53.5      & 58.6   & 8 \\
        Au$_{3}$Ni            & 66.25     & 66.25  & 4 \\
        Au75Ni25    & 53.5      & 66.25   & 8 \\
        Au80Ni20    & 53.5      & 68.8   & 8 \\
        Au99Ni1     & 53.5      & 78.49  & 8 \\
        Au                    & 79.0      & -      & 4 \\
        \bottomrule
    \end{tabular}
\end{table}

Inspection of the EMsoft v5 source code \cite{EMsoft} used in this work (fork commit: 2c071d69) showed that partial site occupancy is treated consistently in the dynamical structure-factor calculation, but inconsistently in two other parts of the simulation chain. Density and average atomic weight are occupancy weighted, whereas the $\bar{Z}$ supplied to the Monte Carlo calculation is obtained from unweighted orbit multiplicities. Therefore, a split Au/Ni site results in a $\bar{Z} = 53.5$ for all intermediate compositions as seen in Tab.~\ref{tab:tab_2}. In addition, the combined master pattern is normalised by the unweighted sum of orbit multiplicities. The latter acts as a global intensity scale and is removed by per-pattern $Z$-score normalisation, but it invalidates direct comparisons of raw mean intensity between structures represented with different numbers of atom records (Tab.~\ref{tab:tab_2}). In contrast, the incorrect $\bar{Z}$ in the Monte Carlo simulation can modify the energy-, depth-, and exit-direction weighting before pattern normalisation and therefore cannot be corrected reliably by image normalisation alone. Consequently, the simulated patterns retain physically meaningful occupancy-dependent diffraction information, including changes in relative reflection strength, while absolute-intensity trends and ordered–disordered comparisons that use different occupancy representations require occupancy-corrected simulations for quantitative interpretation.

\FloatBarrier

\subsubsection{Ordering effect}

The significant increase in number of strong reflections for the ordered state sample, Au$_3$Ni, is consistent with expectation as well. Considering that the atoms have specific atomic sites, in this case the atoms of nickel as fixed in the corners and gold in the face centres, this decreases the translational symmetry. This leads to a change in structure factor and reflections which are forbidden in the unordered state, become possible, the so-called superlattice reflections. Although a higher number of strong reflections were included in the dynamical matrix (compared to the unordered state), which can suggest presence of superlattice reflections, they were not observed through the line profile analysis. Previous reports on the identification of superlattice reflections through line profile analysis on Kikuchi patterns have shown well defined split peaks \cite{Payattuvalappil2025}, rather than what was observed in Fig.~\ref{fig:fig_10}(d).

\subsection{Changes in normalised intensity distribution}

Finally, a difference in normalised intensity distribution was observed as the chemical composition and the degree of order was varied (Fig.~\ref{fig:fig_5}(d)-(f) and \ref{fig:fig_10}(b)). This was particularly highlighted in \{111\} and \{002\} bands, for both cases, and also in some zone axis, especially for the ordered Au$_3$Ni sample. In both cases, this could be attributed to a change in the structure factor. For chemical composition variation, as $\bar{Z}$ increases, it increases the scattering factor, changing the structure factors which can lead to the difference in normalised band intensity distribution observed. A change in structure factor is also expected due to ordering, which can lead to change in band intensity, in addition to superlattice reflections becoming possible. Therefore, this results in the redistribution of normalised intensity observed in Fig.~\ref{fig:fig_10}(b), where some areas exhibits higher normalised scattering intensity (highlighted in red) and others lower (highlighted in blue). This is also noticeable in the line profile measurements, where less prominent higher order peaks are observed. 

\subsection{Considerations for future machine learning models}

Investigating the subtle changes in the Kikuchi patterns as aspects of the material are varied is useful for the development of representation learning based indexing methods for challenging structures, such as ordered phases or phases containing structurally complex defect phases. In this work, the focus was on changes in chemical composition, lattice parameter and ordering. Systematically analysing these pattern changes provides a baseline for selecting the preprocessing procedures and later interpretation of the features learned by the model. Ideally, the ML model should learn the physically meaningful features of the Kikuchi pattern, supporting a more physics-informed approach rather than relying on image-to-image comparison, while also avoiding artifacts or non-physical aspects \cite{Lapuschkin-ml-learn-2019}. In the patterns analysed, such artifacts can arise from the misleading raw intensity values observed for partial site occupancy simulations. This highlights the need for an intensity normalisation step prior to training, while ensuring that the normalisation preserves the other relevant physical descriptors. Additionally, understanding how the Kikuchi pattern changes with chemical composition, lattice parameter, and ordering provides a basis for assessing what the ML model may encode in the latent representation of the patterns. The most relevant descriptors identified were the normalized intensity distribution and band width, which changed as chemical composition and lattice parameter were varied and similarly observed for ordering.

More generally, this result illustrates a particular vulnerability of simulation-based full-pattern methods. Their sensitivity to relative band intensities is what enables dictionary indexing, spherical matching, sublattice discrimination and representation learning to distinguish structures that have almost identical band geometry. The same sensitivity also makes them dependent on the fidelity of the forward model. The present inconsistency does not affect fully occupied structures in this way and is expected to have limited influence on orientation-dominated analyses in which patterns are individually normalised. However, composition-, occupancy-, or order-sensitive analyses can become confounded when a physical class is systematically associated with a different crystallographic encoding, for example when a fully occupied ordered structure is compared with a partially occupied disordered structure. Such comparisons should use occupancy-corrected code, verify that physically equivalent crystal encodings produce equivalent patterns, and, at fixed overall composition, reuse a common Monte Carlo transport calculation across different states of order.

\section{Conclusions}

In this work, a systematic investigation of simulated Kikuchi patterns is presented, and the effects of chemical composition and lattice parameter variations and ordering in the Au-Ni patterns were analysed. The aim was to understand how Kikuchi patterns behave as aspects of the materials are systematically modified, thereby providing a baseline for future implementation of ML models in new indexing methods, particularly for challenging structures such as ordered phases or phases containing structurally complex defect phases. Our findings were consistent with expectations as increasing gold content led to an increase in mean intensity and a higher number of strong reflections were included in the dynamical matrix. Unexpectedly, limitations of EMsoft regarding partial site occupancy weighting led to a sudden four-fold increase in mean intensity from 99\% gold to pure gold. Since raw intensity is a relevant descriptor of a Kikuchi pattern, this could be a concern when implementing ML models for phase or defect state indexing and an intensity normalisation step needs to be considered. The analysis of the isolated effects of lattice parameter and chemical composition showed that the chemical composition contributes predominantly to the mean intensity and strong reflections count, although the normalised intensity distribution was affected by both. When L$1_2$ ordering was introduced, a higher mean intensity and a higher number of strong reflections compared to the unordered state were observed, due to the inclusion of superlattice reflections forbidden in the unordered state. In addition, a redistribution of the normalised intensity was observed. However, the absolute-intensity difference between the ordered and disordered structures was confounded by inconsistent occupancy handling in the Monte Carlo and master pattern normalisation paths and requires a corrected comparison using a common Monte Carlo result. The results revealed differences in normalised intensity distribution, as well as band width, between the analysed patterns, and therefore identified as key descriptors. This study provides a better understanding of the descriptors that an ML model can learn from Kikuchi patterns, incorporating other meaningful features beyond band position into future indexing methods.

\section{Declaration of competing interest}

The authors declare that they have no known competing financial interests or personal relationships that could have appeared to influence the work reported in this paper.

\section{CRediT authorship contribution statement}

\textbf{Camila A. Teixeira}: Writing - original draft, Visualization, Data curation, Investigation, Formal analysis. \textbf{Lukas Berners}: Writing - review \& editing, Formal analysis, Visualization. \textbf{Sandra Korte-Kerzel}: Writing - review \& editing, Formal analysis, Supervision, Project administration, Funding acquisition. \textbf{Ulrich Kerzel}: Writing - review \& editing, Supervision, Funding acquisition, Formal analysis, Software, Conceptualization.

\section{Declaration of generative AI and AI-assisted technologies in the manuscript preparation process}

During the preparation of this work, the authors used RWTHgpt to check spelling and grammar and improve the conciseness and readability of existing text passages. After using the RWTHgpt, the authors reviewed and edited the content as needed and take full responsibility for the content of the published article.

\section{Data availability}

The data used in this work can be made available upon request.

\section{Acknowledgements}

The authors acknowledge the support of the German Research Foundation (DFG) within the Collaborative Research Centre SFB 1394 "Structural and Chemical Atomic Complexity - From Defect Phase Diagrams to Materials Properties" (project ID 409476157), including project groups C02 and A07. The data used in this publication was managed using the framework and metadata scheme provided by project A07 within SFB 1394 (project ID 409476157) funded by the Deutsche Forschungsgemeinschaft (DFG), and using the research data management platform Coscine with storage space granted by the Research Data Storage (RDS) of the DFG and Ministry of Culture and Science of the State of North Rhine-Westphalia (DFG: INST222/1261-1 and MKW: 214-4.06.05.08 - 139057). The computation of master and screen patterns used in this work were carried out with the computing resources granted by the RWTH Aachen University under project rwth1955. The authors express thanks to Alexander Schwedt and Joshoua Spille for providing and metallographically preparing the pure nickel sample used to acquire the experimental patterns used in this work.

\printbibliography

\end{document}